\documentclass[12pt]{article}

\usepackage{amssymb}
\usepackage{amsmath}
\allowdisplaybreaks

\numberwithin{equation}{section}

\newcommand{\ve}[1]{\boldsymbol{#1}}
\newcommand{\half}{\frac{1}{2}}
\newcommand{\Nc}{{\cal N}_{\rm c}}
\newcommand{\cQ}{{\cal Q}}
\newcommand{\cM}{{\cal M}}

\newcommand{\bra}[1]{\langle #1 |}
\newcommand{\ket}[1]{| #1 \rangle}
\newcommand{\bracket}[2]{\langle #1| #2 \rangle}
\newcommand{\del}{\partial}
\newcommand{\nn}{\nonumber}
\newcommand{\Pmatrix}[1]{\begin{pmatrix} #1 \end{pmatrix}}
\newcommand{\Smatrix}[1]{
\left(\begin{smallmatrix} #1 \end{smallmatrix}\right)}
\newcommand{\vL}{\bigr|_L}
\newcommand{\K}{K_1}
\newcommand{\T}{{\rm T}}
\newcommand{\ff}{{\bf f}}
\newcommand{\uu}{{\bf u}}
\newcommand{\ee}{{\bf e}}
\newcommand{\en}[1]{\sqrt{#1} \ee_{#1}}
\newcommand{\dd}{{\bf d}}

\newcommand{\sign}[1]{(-1)^{\frac{#1}{2}}}
\newcommand{\chs}[2]{\begin{pmatrix}#1 \\ #2\end{pmatrix}}

\newcommand{\wtTsT}{\widetilde{T}\,\widetilde{\star}\,\widetilde{T}}

\begin{document}

\baselineskip=17.0pt plus 0.2pt minus 0.1pt

\begin{titlepage}
\title{
\hfill\parbox{4cm}
{\normalsize KUNS-1839\\{\tt hep-th/0305010}}\\
\vspace{1cm}
{\bf Gauge Structure of\\ Vacuum String Field Theory}
}
\author{
Hiroyuki {\sc Hata},
\thanks{{\tt hata@gauge.scphys.kyoto-u.ac.jp}}
\quad
Hisashi {\sc Kogetsu}
\thanks{{\tt kogetsu@gauge.scphys.kyoto-u.ac.jp}}
\quad and \quad
Shunsuke {\sc Teraguchi}
\thanks{{\tt teraguch@gauge.scphys.kyoto-u.ac.jp}}
\\[15pt]
{\it Department of Physics, Kyoto University, Kyoto 606-8502, Japan}
}
\date{\normalsize May, 2003}
\maketitle
\thispagestyle{empty}

\begin{abstract}
We study the gauge structure of vacuum string field theory expanded
around the D-brane solution, namely, the gauge transformation and the
transversality condition of the massless vector fluctuation mode.
We find that the gauge transformation on massless vector field is
induced as an anomaly; an infinity multiplied by an infinitesimal
factor. The infinity comes from the singularity at the edge of the
eigenvalue distribution of the Neumann matrix, while the infinitesimal 
factor from the violation of the equation of motion of the fluctuation
modes due to the regularization for the infinity.
However, the transversality condition cannot be obtained
even if we take into account the anomaly contribution.
\end{abstract}

\end{titlepage}

\section{Introduction}
It has been conjectured that vacuum string field theory (VSFT)
\cite{Rastelli:2000hv,Rastelli:2001jb,Rastelli:2001vb,Rastelli:2001uv}
represents a string field theory expanded around the non-perturbative
vacuum on which open string tachyon has condensed.
The VSFT action is obtained from the action of the cubic string
field theory (CSFT) by replacing the BRST operator $Q_{\rm B}$ with a
purely ghost operator $\cQ$.
In order to check this conjecture, a classical solution of VSFT was
constructed, which is expected to correspond to a D-brane.
There have been many attempts to examine whether this solution has
desired properties as a D-brane. The criteria whether it corresponds
to a D-brane includes
the tension of the solution and string spectrum around it.
In the oscillator formalism, the classical solution is constructed as
a squeezed state \cite{Kostelecky:2000hz,Rastelli:2001jb,Hata:2001sq}
and the mass spectrum around this solution is 
reproduced correctly \cite{Hata:2001sq,Hata:2002pn}.
However, the correct D-brane tension has not been obtained in the
oscillator formalism.\footnote{
See \cite{Okawa:2002pd} for a derivation of the correct D-brane
tension in the boundary conformal field theory approach.}

In addition to the above criteria, the theory around a D-brane must
have the gauge structure 
which is a significant property of string theory.
To see this, let us recall the case of the free SFT
which has the Kato-Ogawa BRST operator $Q_{\rm B}$. 
Taking $b_0b_{-1}\ket{0}\lambda$  as the state of the gauge
transformation, we have 
\begin{equation}
\label{g trans for KO}
Q_{\rm B}(b_0b_{-1}\ket{0}\lambda)
=b_{-1}\ket{0}\square\lambda
+ b_0\alpha^\mu_{-1}\ket{0}\del_\mu \lambda.
\end{equation}
The second term is proportional to the massless vector
state and generates the gauge transformation
$\delta A_\mu=\del_\mu\lambda$ on the massless vector field $A_\mu$.
On the other hand, let us consider the equation of motion for the
massless vector state with polarization $\zeta_\mu$:
\begin{equation}
0=Q_{\rm B}(b_0\alpha_{-1}^\mu\ket{0}\zeta_\mu)
=\alpha^\mu_{-1}\ket{0}\square \zeta_\mu +
b_0c_{-1}\ket{0}\del^\mu \zeta_\mu.
\label{QBb0a-10}
\end{equation}
We obtain the transversality condition $\del_\mu\zeta^\mu=0$ from the
last term, while the first one provides the on-shell condition for
$\zeta_\mu$.
If the classical solution of VSFT represents a D-brane, VSFT
expanded around it must reproduce the gauge transformation and the
transversality condition of the massless vector state (and also of the
massive higher level states).

The purpose of this paper is to investigate whether these kinds of
gauge structure appear in VSFT using its oscillator formulation.
This is in fact a non-trivial problem since a naive analysis around
the classical solution using the oscillator formulation leads to the
conclusion that neither the gauge transformation nor the transversality
condition can arise from VSFT.
We have already seen the difficulty in reproducing the transversality
condition of massless vector state in \cite{Hata:2001sq}, and the
situation is almost the same also for the gauge transformation.
In this paper, we reexamine the gauge structure
concerning the massless vector state in detail and find that important
terms were overlooked in the naive analysis; an infinite factor
multiplied by an infinitesimal one.
The infinite factor originates from the singularity at the edge of the
eigenvalue distribution of the Neumann matrices defining the
$*$-product of string fields \cite{Hata:2001wa,Rastelli:2001hh}.
The same kind of singularity has made
various observables in VSFT non-trivial ones.
On the other hand, the infinitesimal factor is due to the failure of
the equation of motion of fluctuation modes in the regularization
introduced for controlling the infinity \cite{Hata:2002xm}.

In our analysis of the gauge structure, the anomaly like (namely,
$\infty\times 0$) contribution
mentioned above indeed leads to the desired gauge transformation of
the massless vector field in VSFT. 
The equation of motion of the massless vector state 
$\Phi_{\rm v}$, $\cQ_{\rm B}\Phi_{\rm v}=0$ with $\cQ_{\rm B}$ being
the BRST operator of VSFT around the classical solution, also receives
the anomaly correction.
However, contrary to the case of (\ref{QBb0a-10}) which consists of
two independent terms, the on-shell condition and the transversality
condition, $\cQ_{\rm B}\Phi_{\rm v}=0$ is satisfied only by the
on-shell condition. The role of the anomaly term is merely to shift
the gauge of the on-shell condition and it cannot generate the
transversality condition. 
Therefore, our finding in this paper is that the gauge transformation
on the massless vector field is indeed generated as an anomaly, but
the absence of the transversality condition still remains a mystery to
be resolved.
Our analysis in this paper is restricted to the Fock space,
namely, for all the ket equations we are implicitly considering their
inner product with any Fock space bra states.
The extension to larger space including the sliver states
is also our future problem.

Before closing this section we shall give two comments concerning
the gauge structure in VSFT.
First, as seen from (\ref{g trans for KO}) and (\ref{QBb0a-10}), the
ghost sector of states as well as 
the BRST operator plays a vital role in the gauge
structure of the ordinary string field theory.
On the other hand, in VSFT with purely ghost BRST operator, the ghost
and matter sectors are factorized both in the classical solution and
the fluctuation modes. Due to this property, the ghost sector has
played no significant roles in the physics around the D-brane solution.
The gauge structure of VSFT is interesting since this is the first
place where the ghost sector makes an essential contribution.

Our second comment is on another role of gauge transformation in VSFT.
The gauge transformation in VSFT has been used in the
construction of massive and massless states around the D-brane
solution. There, an infinite number of spurious states particular to
VSFT have been gauged away by a special kind of gauge
transformation \cite{Imamura:2002rn,Hata:2002pn}.
This gauge transformation is of completely different type from that
we shall study in this paper.

The organization of the rest of this paper is as follows.
In the next section, we summarize the VSFT action, its classical
solutions and fluctuation modes.
In sec.\ 3, we recapitulate the treatment of the singular behavior
of the infinite dimensional matrices in the oscillator formulation of 
VSFT. This singular behavior plays a crucial role in the following
analysis.
Sec.\ 4  is the main part of this paper, and we investigate the gauge
structure of the VSFT expanded around the D-brane solution.
The final section (sec.\ 5) is devoted to a summary and discussions.
In appendix A--C, we summarize several technical details used in the
text.

\section{VSFT in the oscillator formalism}\label{VSFT action}
In this section, we review the VSFT action, its classical
solution and tachyon wave function as a fluctuation
around the classical solution.
The VSFT action proposed in
\cite{Rastelli:2000hv,Rastelli:2001jb,Rastelli:2001uv} is
\begin{align}
\label{VSFT-action}
S&=-K\left(\half\Psi\cdot\cQ\Psi + \frac{1}{3}\Psi\cdot
(\Psi*\Psi)\right)\nn\\
&=-K\left(\half\int_{b_0,x}\bra{\Psi}\cQ\ket{\Psi}
+\frac{1}{3}\int_{b_0,x}^{(1)}\int_{b_0,x}^{(2)}\int_{b_0,x}^{(3)}
\phantom{}_1\bra{\Psi}_2\bra{\Psi}_3\bracket{\Psi}{V}_{123}
\right),
\end{align}
where $K$ is a constant, and the integration 
$\int^{(r)}_{b_0,x}\equiv\int db^{(r)}\int d^{26}x_r$ is over the zero
modes of the $r$-th string.
The BRST operator $\cQ$ of VSFT consists purely of ghost oscillators,
\begin{equation}
\cQ=c_0+\sum_{n=1}^\infty f_n\bigl(c_n+(-1)^n c_n^\dagger\bigr),
\label{cQ}
\end{equation}
with $f_n$ being a constant.
The VSFT action (\ref{VSFT-action}) has an invariance under the gauge
transformation:
\begin{equation}
\label{gauge-transformation}
\delta_\Lambda \Psi=\cQ\Lambda+\Psi*\Lambda-\Lambda*\Psi.
\end{equation}
The $*$-product in (\ref{VSFT-action}) and
(\ref{gauge-transformation}) is the same as in the
CSFT action and defined through the three string vertex $\ket{V}$,
\begin{align}
\ket{V}_{123}=
&\exp\left\{\sum_{r,s=1}^3\left(-\sum_{n,m\geq 0}\half 
a_n^{(r)\dagger}V_{nm}^{rs}a_m^{(s)\dagger}
+\sum_{n\geq 1,m\geq 0}
c_n^{(r)\dagger}\widetilde V_{nm}^{rs}b_m^{(s)\dagger}
\right)\right\}\ket{0}_{123}\nn\\
&\qquad\times(2\pi)^{26}\delta^{26}(p_1+p_2+p_3),
\end{align}
where $a_0^{(r)}=a_0^{(r)\dagger}=\sqrt{2}\,p_r$ is the center-of-mass 
momentum.
The coefficients $V_{nm}^{rs}$ of the oscillators are infinite
dimensional matrices and are referred to as Neumann coefficients.
It is convenient to introduce the following new matrices and vectors:
\begin{align}
&[M_0]_{mn}=[CV^{r,r}]_{mn},\quad [M_\pm]_{mn}=[CV^{r,r\pm 1}]_{mn},\\
&[\ve{v}_0]_n=[V^{r,r}]_{n,0},\quad 
[\ve{v}_\pm]_n=[V^{r,r\pm 1}]_{n,0},\\
&V_{00}=[V^{r,r}]_{00}=\frac{1}{2}\ln\!\left(\frac{3^3}{2^4}\right),
\end{align}
where $m$ and $n$ run from 1 to infinity, and $C$ is the twist matrix:
\begin{equation}
C_{mn}=\delta_{mn}(-1)^n .
\end{equation}
The Neumann matrices and vectors in the ghost sector are denoted by
adding a tilde to the corresponding one in the matter sector;
$\widetilde M_{0,\pm}$ and $\widetilde{\ve{v}}_{0,\pm}$.
We often use the following combination of the Neumann coefficients:
\begin{align}
M_1&\equiv M_+-M_-,\\
\ve{v}_1&\equiv \ve{v}_+-\ve{v}_-.
\end{align}
Note that $M_1$ and $\ve{v}_1$ are twist-odd, while $M_0$ and
$\ve{v}_0$ are even.
It is known that these matrices satisfy the following
relations\footnote{
Various formulas for the ghost Neumann coefficients are summarized in
appendix \ref{ghost neumann}. }:
\begin{align}
&M_0+M_++M_-=1,\label{3M}\\
&\ve{v}_0+\ve{v}_++\ve{v}_-=0,\\
&[M_0,M_1]=0,\label{commutation of M} \\
&M_1^2=(1-M_0)(1+3M_0), \label{relation of M}\\
&3(1-M_0)\ve{v}_0+M_1\ve{v}_1=0, \\
&3M_1\ve{v}_0+(1+3M_0)\ve{v}_1=0 .\label{M and v}
\end{align}
The Neumann coefficients $M_0,M_1,\ve{v}_0$ and $\ve{v}_1$ can be
represented by a simpler ones $K_1$ and $\uu$
\cite{Rastelli:2001hh,Hata:2002it}:
\begin{align}
M_0&=-\frac{1}{1+2\cosh\!\left(K_1\pi/2\right)},\label{K_1rep}\\
M_1&=\frac{2\sinh\!\left(K_1\pi/2\right)}{
1+2\cosh\!\left(K_1\pi/2\right)}, \\
\ve{v}_0&=-\frac{1}{3}(1+3M_0)\uu,\\
\ve{v}_1&=M_1\uu,
\end{align}
where the matrix $K_1$ and the vector $\uu$ are given by
\begin{align}
[K_1]_{mn}&=-\sqrt{n(n+1)}\,\delta_{m,n+1}-\sqrt{n(n-1)}\,\delta_{m,n-1}, 
\label{K_1}\\
\label{u}
[\uu]_n&=
\frac{1}{\sqrt{n}} \cos\!\left(\frac{n\pi}{2}\right)=
\begin{cases}
\displaystyle\frac{\sign{n}}{\sqrt{n}}
&(n: \mbox{even})
\\
0
&(n: \mbox{odd})
\end{cases}
.
\end{align}
Here, $\K$ is the matrix representation of the Virasoro algebra
$\K=L_1+L_{-1}$ and is twist-odd, $C\K C=-\K$.
The eigenvalue problem of the matrix $K_1$ has been solved
in \cite{Rastelli:2001hh}.
The zero eigenvector $\ff^{(0)}$ of $K_1$ satisfying $K_1\ff^{(0)}=0$
will be important in later discussions. It is a twist-odd vector and
explicitly given by
\begin{align}
\label{fzero}
[\ff^{(0)}]_n=
\frac{1}{\sqrt{n}} \sin\!\left(\frac{n\pi}{2}\right)=
\begin{cases}
0
&(n: \mbox{even})
\\
\displaystyle\frac{\sign{n-1}}{\sqrt{n}} 
&(n: \mbox{odd})
\end{cases}
.
\end{align}

A classical solution of VSFT which satisfies
\begin{equation}
\label{eq of motion}
\cQ \Psi_{c}+\Psi_{c}*\Psi_{c}=0,
\end{equation}
and is expected to represent a D25-brane has been found 
\cite{Kostelecky:2000hz,Rastelli:2001jb,Hata:2001sq}:
\begin{equation}
\label{psi_c}
\ket{\Psi_{\rm c}}=\Nc b_0\exp
\Bigl(-\half \ve{a}^\dagger\cdot CT \ve{a}^\dagger
-\ve{b}^\dagger\cdot C\widetilde T \ve{c}^\dagger\Bigr)\ket{0},
\end{equation}
where $\Nc$ is a normalization factor and $T$ is an infinite
dimensional twist-even matrix. The matrix $T$ is a solution to
\begin{align}
T&=T\star T, \label{equation of T}
\end{align}
where $T\star T$ is defined by
\begin{equation}
T\star T\equiv M_0+(M_+,M_-)(1-T\cM)^{-1}T\left(
\begin{matrix}
M_-\\
M_+\\
\end{matrix}
\right) \label{T*T},
\end{equation}
with
\begin{equation}
\cM=
\begin{pmatrix}
M_0&M_+\\
M_-&M_0
\end{pmatrix}.
\end{equation}
Eq.\ (\ref{equation of T}) can be solved by using the relations
(\ref{3M})--(\ref{M and v}) and assuming the commutativity among $T$
and $M_{0,\pm}$.
We take the following one as a solution of (\ref{equation of T}):
\begin{equation}
\label{T}
T=\frac{1}{2M_0}\left(1+M_0-\sqrt{(1-M_0)(1+3M_0)}\right).
\end{equation}
The equation of motion (\ref{eq of motion}) also fixes the coefficient 
$\ve{f}$ in $\cQ$ (\ref{cQ}) as follows
\cite{Hata:2001sq,Kishimoto:2001ac,Okuyama:2002yr}: 
\begin{equation}
\label{f}
\ve{f}=(1-\widetilde T)^{-1}\left[
\widetilde{\ve{v}}_0+(\widetilde{M}_+,\widetilde{M}_-)
(1-\widetilde T\widetilde\cM)^{-1}\widetilde T
\chs{\widetilde{\ve{v}}_+}{\widetilde{\ve{v}}_-}
\right] ,
\end{equation}
where $\widetilde T$ is given by (\ref{T}) with $M_0$ replaced by
$\widetilde M_0$ (see appendix A for details).

Now, we expand the string field $\Psi$ around the classical solution
${\Psi_{\rm c}}$:
\begin{equation}
\Psi = \Psi_{\rm c}+\Phi.
\end{equation}
The linearized equation of motion for the fluctuation $\Phi$
is\footnote{
The action of the BRST operator $\cQ_{\rm B}$ around $\Psi_{\rm c}$ on
a generic string field $A$ is defined by 
$\cQ_{\rm B}A\equiv\cQ A+\Psi_{\rm c}*A-(-)^{|A|} A*\Psi_{\rm c}$ with 
$|A|=0$ ($1$) if $A$ is Grassmann-even (-odd).}
\begin{equation}
\label{fluc_on-shell}
\cQ_{\rm B}{\Phi}\equiv
\cQ{\Phi}+{\Psi_{\rm c}*\Phi}+{\Phi*\Psi_{\rm c}}=0.
\end{equation}
We take the following state $\ket{\Phi_{\rm t}}$ as the tachyon
fluctuation mode \cite{Hata:2001sq}:
\begin{equation} 
\label{tachyon}
\ket{\Phi_{\rm t}}=\frac{1}{\Nc}\exp\Bigl(
-\sum_{n\geq 1}t_n a_n^\dagger a_0+ip\cdot\hat x\Bigr)
\ket{\Psi_{\rm c}}.
\end{equation}
The linearized equation of motion (\ref{fluc_on-shell}) for the
tachyon state (\ref{tachyon}) determines the condition for the infinite
dimensional vector $\ve{t}$ as
\begin{equation}
\label{eq of t}
\ve{t}=T\star\ve{t},
\end{equation}
with the vector $T\star\ve{t}$ defined by
\begin{equation}
\label{T*t}
T\star \ve{t}\equiv\ve{v}_0-\ve{v}_++(M_+,M_-)(1-T\cM)^{-1}
\left[T\chs{\ve{v}_+-\ve{v}_-}{\ve{v}_--\ve{v}_0}
+\chs{0}{\ve{t}}
\right].
\end{equation}
Using the relations (\ref{3M})--(\ref{M and v}) and the twist-even
property of $\ve{t}$,
\begin{equation}
C\ve{t}=\ve{t},
\end{equation}
(\ref{eq of t}) is solved to give
\begin{equation}
\label{tachyon wave function}
\ve{t}=3(1+T)(1+3M_0)^{-1}\ve{v}_0.
\end{equation}
The linearized equation of motion (\ref{fluc_on-shell}) also
determines the mass squared of the state (\ref{tachyon}):
\begin{equation}
(\mbox{tachyon mass})^2=-1.
\end{equation}
This is the correct mass squared for the open string tachyon
(we are taking the convention of $\alpha'=1$).

\section{Singular behavior around the zero-mode of $\K$}
In the calculation of the tachyon mass and other quantities in the
oscillator formalism of VSFT, we are often faced with the phenomenon
where the zero-mode of $\K$ plays an important role.
This is the case also in our analysis of the gauge structure in VSFT.
Here, we briefly review this phenomenon.
See \cite{Hata:2001wa,Hata:2002it} for more details.

Various quantities in VSFT are expressed in terms of Neumann
coefficients.
For example, the tachyon mass squared is given by $-\ln 2/G$ with
\begin{equation}
G=\frac{9\sqrt{3}}{32}\,\ve{v}_1^{\rm T}\biggl(
\frac{1-M_0}{\sqrt{1+3M_0}}-M_1\frac{1}{(1+3M_0)^{3/2}}M_1
\biggr)\ve{v}_1.
\label{eq:Gdeform}
\end{equation}
This quantity is exactly equal to zero if we naively use the
non-linear relations (\ref{commutation of M})--(\ref{M and v}) for
the Neumann coefficients.
However, the expression (\ref{eq:Gdeform}) is in fact 
{\em indefinite} because the matrix $M_0$ has the eigenvalue $-1/3$. 
The corresponding eigenvector is the zero-mode $\ff^{(0)}$ of $K_1$.

In the correct treatment, we must regularize $G$ properly, and
we use the finite-size matrix regularization following
\cite{Hata:2001wa,Hata:2002it,Hata:2002xm}.
In this regularization, we truncate the infinite dimensional matrices
$M_0,M_1$ into $L\times L$ ones.
After this regularization, the non-linear relations 
(\ref{commutation of M})--(\ref{M and v}) no longer hold.
Because only the parts around the singularity at $K_1=0$ make finite
contributions to the quantity $G$, it is sufficient to Laurent-expand
around $K_1=0$: 
\begin{align}
(1+3M_0)|_L&\simeq\frac{\pi^2}{12}K_1^2|_L,\\
M_1|_L&\simeq\frac{\pi}{3}K_1|_L,\\ 
\ve{v}_1|_L&\simeq\frac{\pi}{3}(K_1\uu)|_L .
\label{v1simeqK1u}
\end{align}
Note that the truncation of the squared matrix, $K_1^2|_L$, is
different from the square of the truncated matrix, $(K_1|_L)^2$.
With these manipulations, $G$ is expressed as follows:
\begin{equation}
G=\frac{\pi}{4}(\uu^TK_1)|_L\left(\frac{1}{\sqrt{K_1^2|_L}}-K_1|_L
\left(\frac{1}{\sqrt{K_1^2|_L}}\right)^3K_1|_L\right)(K_1\uu)|_L.
\end{equation}
Taking the limit $L\rightarrow \infty$ after calculation with the
finite-size matrices, we get the finite and expected value $G=\ln 2$.
The reason why the quantity $G$ vanishes upon use of the non-linear
relations of the Neumann coefficients is reduced to the degeneracy of
the eigenvalues of $M_0$ between twist-odd sector and twist-even one
\cite{Hata:2001wa}.
Therefore, the phenomenon that a quantity such as $G$ which naively
vanishes due to this degeneracy acquires a non-zero value is called
twist anomaly.
 
As we have seen in this section,
the singular behavior around the zero-mode of $\K$ makes 
VSFT around the classical solution non-trivial.
Besides the above example, this zero-mode plays a key role in
constructing the higher excitation modes around the classical solution
\cite{Imamura:2002rn,Hata:2002pn}. 
In the next section, we shall see that a similar phenomenon happens 
in the analysis of the gauge structure of VSFT.

\section{Gauge structure of VSFT}
If the classical solution (\ref{psi_c}) of VSFT  describes
a D25-brane, VSFT expanded around it has to reproduce the
ordinary open string theory. In particular, it must 
have the gauge structure of string theory. 
This section is devoted to the analysis of the gauge structure of VSFT
expanded around the classical solution and is the main part of this
paper.
Here, we concentrate on the gauge structure of the massless vector
field mentioned in the introduction;
the gauge transformation $\delta A_\mu=\del_\mu\lambda$ and the
transversality condition $\del_\mu A^\mu=0$.

First, let us study the gauge transformation on the massless vector
field.
The massless vector state around the classical solution in VSFT is
constructed in \cite{Imamura:2002rn} and is given by
\begin{equation}
\ket{\Phi_{\rm v}}=
\zeta^\mu\bar{\ff}^{(0)}\cdot\ve{a}_\mu^{\dagger}
\ket{\Phi_{\rm t}},
\label{massless vector state}
\end{equation}
where $\zeta^\mu$ is a polarization vector and $\bar{\ff}^{(0)}$ is
the normalized zero eigenvector of $K_1$.
The VSFT gauge transformation (\ref{gauge-transformation}) is
rewritten for the fluctuation $\Phi$ around $\Psi_{\rm c}$ as 
\begin{equation}
\delta_\Lambda \Phi=\cQ_{\rm B}\Lambda+\Phi*\Lambda-\Lambda*\Phi.
\end{equation}
Here, we are interested in the inhomogeneous part:
\begin{equation}
\label{Q_B lambda}
\cQ_{\rm B}\Lambda=
\cQ\Lambda+\Psi_{\rm c}*\Lambda-\Lambda*\Psi_{\rm c}.
\end{equation}
As a candidate $\Lambda$ which induces the massless vector gauge
transformation, we take
\begin{equation}
\label{g_para}
\ket{\Lambda}=\ve{h}\cdot\ve{b}^{\dagger}\ket{\Phi_{\rm t}}.
\end{equation}
Since the massless vector state (\ref{massless vector state}) is
twist-odd, the gauge transformation (\ref{Q_B lambda}) and hence
$\Lambda$ itself must be so. Therefore the vector $\ve{h}$ in
(\ref{g_para}) must be a twist-odd vector satisfying
$C\ve{h}=-\ve{h}$.

The term $\ket{\Psi_{\rm c}*\Lambda}$ in (\ref{Q_B lambda}) is
explicitly given in terms of the Neumann coefficients as follows:
\begin{align}
\label{lambda*psi}
&\ket{\Psi_{\rm c}*\Lambda}=
-2^{-p^2}\left\{(0,\ve{h})(1-\widetilde\cM\widetilde T)^{-1}
\chs{\widetilde{\ve{v}}_+}{\widetilde{\ve{v}}_-}\right.\nn\\
&\hspace{4cm}+\left.\Bigl[(0,\ve{h})(1-\widetilde\cM\widetilde T)^{-1}
\chs{\widetilde{M}_-}{\widetilde{M}_+}\ve{b}^\dagger\Bigl]
\Bigl[c_0+\ve{c}^\dagger(1-\widetilde T)\ve{f}\Bigr]\right\}\nn\\
&\qquad
\times b_0\exp\left\{-(T\star\ve{t})\cdot\ve{a}^\dagger a_0
+ip\cdot\hat x
-\frac{1}{2}\ve{a}^\dagger\cdot C(T\star T)\ve{a}^\dagger
-\ve{b}^\dagger\cdot C(\wtTsT)\ve{c}^\dagger
\right\}
\ket{0},
\end{align}
where $T\star\ve{t}$ and $T\star T$ are defined by (\ref{T*t}) and
(\ref{T*T}), respectively. The other term 
$-\ket{\Lambda*\Psi_{\rm c}}$ in (\ref{Q_B lambda}) is given by
(\ref{lambda*psi}) with $(0,\ve{h})$ and $T\star \ve{t}$ replaced by
$(\ve{h},0)$ and $C(T\star\ve{t})$, respectively.
Namely, $\ket{\Lambda*\Psi_{\rm c}}$ is the twist transform of
$\ket{\Psi_{\rm c}*\Lambda}$.\footnote{
The twist transformation property of the $*$-product of generic
string fields $A$ and $B$ in our convention is
$(\Omega A)*(\Omega B)=(-1)^{|A||B|}\Omega(B*A)$, where $\Omega$ is
the twist transformation operator acting on string fields.}
Note that the state in the last line of (\ref{lambda*psi}) is the
tachyon state (\ref{tachyon}) with the replacements 
$(\ve{t}, T, \widetilde T)\to
(T\star\ve{t}, T\star T, \wtTsT)$.

In deriving (\ref{lambda*psi}), we have not used any one of the
non-linear relations (\ref{3M})--(\ref{M and v}) which are potentially
invalid in the finite-size regularization, except at $2^{-p^2}$ which
has been obtained by applying the calculation of sec.\ 3 to its
original expression $e^{-Gp^2}$.
Although we need to carefully treat the Neumann coefficients
as described in the previous section, let us first simplify
(\ref{lambda*psi}) by naive manipulations.
Using the non-linear relations (\ref{3M})--(\ref{M and v}) and the
equations (\ref{equation of T}) and (\ref{eq of t}), we have
\begin{equation}
\label{Q_B lambda naive}
\cQ_{\rm B}\ket{\Lambda}|_{\rm naive}
=(1-2^{-p^2})
\ve{h}\cdot\ve{b}^\dagger
\left(c_0+[\ve{c}^\dagger(1-\widetilde T)\ve{f}]\right)
\ket{\Phi_{\rm t}}.
\end{equation}
The part of (\ref{Q_B lambda naive}) multiplied by
$1$ of $(1-2^{-p^2})$ has come from $\cQ\ket{\Lambda}$.
Since the state (\ref{Q_B lambda naive}) does not contain the massless
vector state (\ref{massless vector state}), we cannot obtain the
gauge transformation of the massless vector field in this naive
treatment.

However, a careful treatment by taking into account the singularity at 
$K_1=0$ mentioned in sec.\ 3 will show that $\cQ_{\rm B}\Lambda$
(\ref{Q_B lambda}) does contain an additional term proportional to the
massless vector state.
To see this, let us first consider the following term in 
(\ref{lambda*psi}):
\begin{equation}
D=(0,\ve{h})
(1-\widetilde\cM\widetilde T)^{-1}
\chs{\widetilde{\ve{v}}_+}{\widetilde{\ve{v}}_-}
=-\half\ve{h}\cdot(1,-1)
(1-\widetilde\cM\widetilde T)^{-1}
\chs{\widetilde{\ve{v}}_+}{\widetilde{\ve{v}}_-} ,
\label{D}
\end{equation}
where we have used the fact that $\ve{h}$ is twist-odd in obtaining
the last expression. The corresponding term 
$(\ve{h},0)(1-\widetilde\cM\widetilde T)^{-1}
\chs{\widetilde{\ve{v}}_+}{\widetilde{\ve{v}}_-}$
in 
$\ket{\Lambda*\Psi_{\rm c}}$ is equal to $-D$.
In the naive expression (\ref{Q_B lambda naive}), these two $D$ terms
have cancelled each other out since they are multiplied by the
twist-even state $\ket{\Phi_{\rm t}}$.
Using the non-linear relations for the Neumann coefficients,
$D$ is calculated to give
\begin{equation}
D=-\half
\ve{h}\cdot[(1-\widetilde{M}_0)(1+\widetilde{T})]^{-1}(1-\widetilde{T})
\widetilde{\ve{v}}_1 .
\label{Ddiv}
\end{equation}
The ghost Neumann  matrix $\widetilde{M}_0$ has an
eigenvector with eigenvalue $1$, which corresponds to the zero-mode
of $\K$ (see appendix \ref{ghost neumann}).
Therefore, this factor $D$ is in general ill-defined and needs to be
regularized. 
As is shown in appendix \ref{calc of D}, $D$ is of the order of
$\sqrt{\ln L}$ in the finite-size regularization,\footnote{
As seen from (\ref{DeqsqrtlnL/n}), the finite factor in $D$ multiplying
$\sqrt{\ln L}$ depends on the choice of $\bar{\ve{p}}_{n-\half}$
as the finite-size version of $\bar\ff^{(0)}$.
In this section, we adopt $n=1$.
}
\begin{equation}
D\simeq -\frac{1}{\pi}\sqrt{\ln L} ,
\label{DsimeqsqrtlnL}
\end{equation}
if we choose as the vector $\ve{h}$ the following one,
\begin{equation}
\label{h}
\ve{h}=E^{-1}\bar\ff^{(0)},
\end{equation}
with 
\begin{equation}
\label{E}
E_{mn}=\delta_{mn}\sqrt{n}.
\end{equation}
For a generic twist-odd vector $\ve{h}$ which does not contain the
zero-mode $\ff^{(0)}$, the factor $D$ is less divergent than
$\sqrt{\ln L}$.

Recall that in deriving the naive expression (\ref{Q_B lambda naive}),
the factor $D$ has been canceled out between
$\ket{\Psi_{\rm c}*\Lambda}$ and $\ket{\Lambda*\Psi_{\rm c}}$.
However, since $D$ is divergent for $\ve{h}$ of (\ref{h}), it can
couple with some infinitesimal twist-odd quantity to produce finite
effects which were overlooked in deriving (\ref{Q_B lambda naive}).
In (\ref{Q_B lambda naive}) we have replaced $T\star\ve{t}$ in the
exponent of (\ref{lambda*psi}) with $\ve{t}$ by using
$\ve{t}=T\star\ve{t}$ (\ref{eq of t}).
However, as was pointed out in \cite{Hata:2002xm}, (\ref{eq of t}) is
not exactly satisfied by the solution (\ref{tachyon wave function}) 
in the finite-size regularization.
In fact, the infinitesimal twist-odd part exists in
$T\star\ve{t}$ though $\ve{t}$ is twist-even.
This infinitesimal twist-odd part multiplied by $D$ can produce a
finite contribution to $\cQ_B\Lambda$.
Besides the twist-odd part of $T\star\ve{t}$, there are other
infinitesimal differences between the exponent of (\ref{lambda*psi})
and that of $\ket{\Phi_{\rm t}}$. For example, there is an
infinitesimal difference between $T\star T$ and $T$  in this
regularization. But these differences are all twist-even and do not
contribute to $\cQ_{\rm B}\Lambda$.

Regularizing the Neumann coefficients to $L\times L$ ones and
expanding them in $K_1$,
the twist-odd part of $T\star\ve{t}$ is given by\footnote{
The first expression of (\ref{T*t odd}) is due to (3.13) of
\cite{Hata:2002xm}, $\Delta T\simeq (\pi/2)\sqrt{K_1^2}$,
$\ve{t}\simeq -(3/2)\bigl(1/\sqrt{K_1^2}\bigr)K_1\ve{v}_1$
and (\ref{v1simeqK1u}).
}
\begin{align}
\label{T*t odd}
\frac{1-C}{2}\,T\star\ve{t}\bigg|_L
&\simeq-\frac{3}{4}
\left(1-K_1|_L\frac{1}{K_1^2|_L}K_1|_L\right)
\frac{\pi}{3}(K_1\uu)|_L\nn\\
&=-\frac{\pi}{4}\left(
\bar{\ff}^{(0)}\bigr|_L\otimes\bar{\ff}^{(0)}\bigr|_L\right)
(K_1\uu)|_L\nn\\
&=-\frac{\pi}{4}\sqrt{\frac{2}{\ln L}}\,\bar{\ff}^{(0)}\bigr|_L,
\end{align}
where we have used the equality (\ref{ff})
and the fact that $(\K\uu)\vL=(1,0,0,\cdots,0)$.
One might think that it is not clear whether the leading contribution
of the twist-odd part of $T\star\ve{t}$ is given by the expansion in
$\K$.
However, this is justified by considering the inner product of 
$\left[(1-C)/2\right]T\star\ve{t}|_L$ with a suitable basis of the
vector space,
and carrying out the same argument as in sec.\ 3 for $G$.
The infinitesimal factor in (\ref{T*t odd}) and the divergent one $D$
appear in $\ket{\Psi_{\rm c}*\Lambda}-\ket{\Lambda*\Psi_{\rm c}}$
in the following way to give a finite contribution:
\begin{equation}
De^{\varepsilon}-De^{-\varepsilon}
\simeq 2D\varepsilon
\label{De}
\end{equation}
with
\begin{equation}
\varepsilon= -\frac{1-C}{2}(T\star\ve{t})\cdot\ve{a}^\dagger a_0 .
\end{equation}
Note that the $\sqrt{\ln L}$ factors are canceled in $D\varepsilon$ on
the r.h.s of (\ref{De}).
Including this new term, (\ref{Q_B lambda}) is now given by
\begin{equation}
\cQ_{\rm B}\ket{\Lambda}=
(1-2^{-p^2})
(\bar{{\ff}}^{(0)}\cdot E^{-1}\ve{b}^\dagger)
\left(c_0+[\ve{c}^\dagger(1-\widetilde T)\ve{f}]\right)
\ket{\Phi_{\rm t}}
+2^{-p^2}p^\mu\,\bar{\ff}^{(0)}\cdot\ve{a}^\dagger_\mu
\ket{\Phi_{\rm t}} .
\label{Q_B lambda full}
\end{equation}
The last term of (\ref{Q_B lambda full}), which is the massless vector
state with the polarization proportional to the center-of-mass
momentum $p^\mu$, does generate the desired gauge transformation on
the massless vector field. 
Recall that this term has emerged from $0\times\infty$ and hence is a
kind of anomaly.

Next let us consider whether the transversality condition can arise
by a similar analysis.
Corresponding to (\ref{lambda*psi}) for $\cQ_{\rm B}\ket{\Lambda}$,
the term $\ket{\Psi_{\rm c}*\Phi_{\rm v}}$
in $\cQ_{\rm B}\ket{\Phi_{\rm v}}$ for the vector state
$\ket{\Phi_{\rm v}}=\ve{\zeta}^\mu\cdot\ve{a}_\mu^\dagger
\ket{\Phi_{\rm t}}$ is given by
\begin{align}
&\ket{\Psi_{\rm c}*\Phi_{\rm v}}=
-2^{-p^2}\Biggl\{
\ve{a}_\mu^\dagger\cdot\left(M_+,M_-\right)\left(1-T\cM\right)^{-1}
\Pmatrix{\ve{\zeta}^\mu \\ 0}
\nn\\
&\quad
+\Bigl[(\ve{t},0)\cM +(\ve{v}_- -\ve{v}_0,\ve{v}_+ -\ve{v}_-)
\Bigr]\left(1-T\cM\right)^{-1}\Pmatrix{\ve{\zeta}^\mu \\ 0}
a_0^\mu\Biggr\}
\left(c_0+\ve{c}^\dagger(1-\widetilde T)\ve{f}\right)
\nn\\
&\quad
\times b_0\exp\left\{-(T\star\ve{t})\cdot\ve{a}^\dagger a_0
+ip\cdot\hat x
-\frac{1}{2}\ve{a}^\dagger\cdot C(T\star T)\ve{a}^\dagger
-\ve{b}^\dagger\cdot C(\wtTsT)\ve{c}^\dagger
\right\}\ket{0} .
\label{Psic*Phiv}
\end{align}
For the genuine massless vector state with polarization $\zeta^\mu$,
(\ref{massless vector state}),
we have $\ve{\zeta}^\mu=\zeta^\mu\bar{\ff}^{(0)}$.
Here for the moment, we treat $\ve{\zeta}^\mu$ as a generic twist-odd
vector.
First, the naive calculation gives
\begin{equation}
\cQ_{\rm B}\ket{\Phi_{\rm v}}|_{\rm naive}=(1-2^{-p^2})
\left\{\ve{\zeta}^\mu\cdot\ve{a}_\mu^\dagger
\left(c_0+[\ve{c}^\dagger(1-\widetilde T)\ve{f}]\right)\right\}
\ket{\Phi_{\rm t}},
\end{equation}
implying that the linearized equation of motion of $\Phi_{\rm v}$
is satisfied by $p^2=0$ for an arbitrary $\ve{\zeta}^\mu$
\cite{Hata:2001sq}.
However, corresponding to $D$ (\ref{Ddiv}) for the gauge
transformation, there exists a dangerous factor also in the present
case. It is
\begin{equation}
\Bigl[(\ve{t},0)\cM +(\ve{v}_- -\ve{v}_0,\ve{v}_+ -\ve{v}_-)
\Bigr]\left(1-T\cM\right)^{-1}\Pmatrix{\ve{\zeta}^\mu \\ 0}
=-\half\ve{v}_1\cdot(1-T)\left[
(1-M_0)(1+T)\right]^{-1}\ve{\zeta}^\mu,
\label{DforPhiv}
\end{equation}
where the second expression has been obtained by using the
non-linear relations for the Neumann coefficients and that
$\ve{\zeta}^\mu$ is twist-odd. Since we have $T=-1$ at $K_1=0$, this
quantity (\ref{DforPhiv}) with
$\ve{\zeta}^\mu=\zeta^\mu\bar{\ff}^{(0)}$ is $\sqrt{\ln L}$ divergent
in the finite-size regularization in the same manner as for $D$.
In fact, the singular part of (\ref{DforPhiv}) is the same as that of
$D$.
Then, taking into account this divergent nature of (\ref{DforPhiv})
and the infinitesimal twist-odd part of $T\star \ve{t}$, we find that 
$Q_{\rm B}\ket{\Phi_{\rm v}}$ with the anomaly contribution included
is given by
\begin{equation}
\cQ_{\rm B}\ket{\Phi_{\rm v}}
=\zeta^\mu\left[(1-2^{-p^2})\delta_\mu^\nu
+\sqrt{2}\,2^{-p^2} p_\mu
p^\nu\right]
\bar{\ff}^{(0)}\cdot\ve{a}^\dagger_\nu
\left(c_0+[\ve{c}^\dagger(1-\widetilde T)\ve{f}]\right)
\ket{\Phi_{\rm t}} .
\label{cQBPhiv}
\end{equation}
Eq.\ (\ref{cQBPhiv}) implies that $\cQ_{\rm B}\Phi_{\rm v}=0$ is
satisfied if only one equation for the polarization,
\begin{equation}
\left(\Bigl(1-2^{-p^2}\Bigr)\eta_{\mu\nu}
+\sqrt{2}\,2^{-p^2} p_\mu p_\nu\right)\zeta^\nu=0 ,
\label{eqforzeta}
\end{equation}
is satisfied. The effect of the anomaly is merely to add the $p_\mu
p_\nu$ term to the naive mass-shell condition
$\bigl(1-2^{-p^2}\bigr)\zeta^\mu=0$,
namely, to change the gauge of the equation.
Therefore, even if we take into account the anomaly contribution, the
transversality condition for $\zeta^\mu$ cannot be obtained.\footnote{
As an example of $\zeta^\mu$ which satisfies (\ref{eqforzeta}) but
is not subject to the transversality condition, we have the dipole
solution,
$$
\zeta^\mu=p^\mu\delta'(p^2)
+\left[(1+\ln 2/\sqrt{2})/(p\cdot\bar p)\right]\bar p^\mu\delta(p^2),
$$
multiplied by a regular function of $p$.
Here, $\bar p^\mu$ is defined by $\bar p^\mu=(p,0,\cdots,0,-p)$ in the 
frame where $p^\mu$ is given by $p^\mu=(p,0,\cdots,0,p)$.
For this $\zeta^\mu$, we have
$p_\mu\zeta^\mu=(\ln 2/\sqrt{2})\delta(p^2)$. 
We would like to thank T.\ Kugo for discussions on this point.
}

\section{Summary}
In this paper, we have investigated the gauge structure of VSFT
expanded around the classical solution which is expected to represent
a D25-brane.
We obtained the correct gauge transformation for the massless vector
state as a kind of anomaly.
The infinite factor from the ghost part and the infinitesimal one
of the matter part cooperate to make a finite contribution to the
gauge transformation.
Unfortunately, the transversality condition cannot be gained even if 
we take into account the anomaly.

The failure in obtaining the transversality condition is a serious
problem. However, it could give a hint to the resolution of other
problems in the oscillator formulation of VSFT, in particular, the
problem of obtaining the correct D-brane tension
\cite{Hata:2001wa,Rastelli:2001wk,Hata:2002it,Hata:2002xm}.
The discussion in this paper is limited to the massless vector state.
It is also an interesting problem to extend our analysis to the higher 
level fluctuation states \cite{Hata:2002pn}.

\section*{Acknowledgments}
We would like to thank T.\ Kugo for valuable discussions and
comments.
The work of H.\,H. was supported in part by a
Grant-in-Aid for Scientific Research from Ministry of Education,
Culture, Sports, Science, and Technology (\#12640264 and \#15540268).

\appendix
\section{Ghost Neumann coefficients}\label{ghost neumann}
We summarize the properties of the ghost Neumann coefficients which
are used in the calculations in this paper.
The ghost Neumann coefficients satisfy the following relations which 
correspond to (\ref{3M})--(\ref{M and v}) for the matter ones:
\begin{align}
&\widetilde M_0+\widetilde M_++\widetilde M_-=1,\\
&\widetilde {\ve{v}}_0+\widetilde {\ve{v}}_+
+\widetilde {\ve{v}}_-=0,\\
&[\widetilde M_0,\widetilde M_1]=0,\\
&\widetilde M_1^2=(1-\widetilde M_0)(1+3\widetilde M_0), \\
&(1+3\widetilde M_0)\widetilde{\ve{v}}_0-\widetilde
M_1\widetilde{\ve{v}}_1=0,\\
&\widetilde M_1\widetilde{\ve{v}}_0
-(1-\widetilde M_0)\widetilde{\ve{v}}_1=0. 
\end{align}
The matrix $\widetilde T$ in (\ref{psi_c}) is a solution to 
$\widetilde T=\wtTsT$ with $\widetilde\star$
defined by (\ref{T*T}) with all the matrices replaced by tilded ones.
It is given by
\begin{equation}
\widetilde T=\frac{1}{2\widetilde M_0}\left(
1+\widetilde M_0-\sqrt{(1-\widetilde M_0)(1+3\widetilde M_0)}\right).
\label{wtT}
\end{equation}
The eigenvalue distribution of $\widetilde M_0$ is in the range $[0,1]$,
and $\widetilde M_0=0$ ($1$) corresponds to $\widetilde T=0$ ($1$).

The ghost Neumann coefficients are related to the matter ones by
\cite{Gross:1986ia}
\begin{align}
\widetilde{M_0}&=-E\frac{M_0}{1+2M_0}E^{-1},\\
\widetilde{M_1}&=E\frac{M_1}{1+2M_0}E^{-1},
\end{align}
where the matrix $E$ is defined by (\ref{E}).
Using these relations, the ghost Neumann coefficients can be
expressed in terms of $E$, $K_1$ (\ref{K_1}) and
$\uu$ (\ref{u}):
\begin{align}
\widetilde M_0&=E\frac{1}{2\cosh (K_1\pi/2)-1}E^{-1},\\
\widetilde M_1&=E\frac{2\sinh (K_1\pi/2)}{
2\cosh (K_1\pi/2)-1}E^{-1},\\
\widetilde{\ve{v}}_0&=(1-\widetilde M_0)E\uu,\\
\widetilde{\ve{v}}_1&=\widetilde M_1E\uu.
\end{align}
Note that $K_1=0$ corresponds to $\widetilde M_0=1$.
Laurent-expansion around $K_1=0$ in the finite-size regularization
gives
\begin{align}
(1-\widetilde{M}_0)|_L&\simeq \frac{\pi^2}{4}E K_1^2|_L E^{-1},\\
\widetilde{M}_1|_L&\simeq\pi E\K|_L E^{-1},\\ 
\widetilde{\ve{v}}_1|_L&\simeq \pi E(K_1\uu)|_L ,
\\
1-\widetilde{T}(\widetilde{M}_1|_L)
&\simeq \frac{\pi}{2} E\sqrt{K_1^2|_L}E^{-1} ,
\label{1-wtTsimeq}
\end{align}
where $\widetilde{T}(\widetilde{M}_1|_L)$ in (\ref{1-wtTsimeq}) is
$\widetilde{T}$ (\ref{wtT}) with $\widetilde{M}_0$ replaced by the
cutoff one $\widetilde{M}_0|_L$.

\section{Calculation of the divergent factor $D$}\label{calc of D}
In this appendix, we calculate the divergent factor $D$ (\ref{Ddiv}).
Using the formulas in appendix \ref{ghost neumann}, $D$ is expanded
around $\K=0$ as
\begin{equation}
\label{D of K}
D\simeq-\half\ve{h}\cdot
E\frac{1}{\sqrt{\K^2\vL}}(\K\uu)\vL,
\end{equation}
where we have kept only the most singular term.
As shown in \cite{Hata:2002it},
$1/\sqrt{\K^2\vL}$ has the following representation:
\begin{equation}
\frac{1}{\sqrt{\K^2\vL}}=
\Pmatrix{
P_{\rm H}\Lambda_{\rm H}^{-1}P_{\rm H}^T & 0 \\
0 & Q_{\rm I}\Lambda_{\rm I}^{-1}Q_{\rm I}^T \\
}.
\end{equation}
Here, we have changed the rows and columns of the matrix and collected
the elements with odd-odd indices in the upper-left block, and vice
versa. The matrix $P_{\rm{H}}$ ($Q_{\rm I}$) is constructed from the
normalized twist-odd (even) eigenvectors
$\bar{\ve{p}}_{n-\frac{1}{2}}$ ($\bar{\ve{q}}_{n}$) 
of $\K^2\vL$ and $\Lambda_{\rm H}$ ($\Lambda_{\rm I}$) is the diagonal
matrices consisting of the corresponding eigenvalues 
$\kappa_{n-\frac{1}{2}}$ ($\kappa_{n}$):
\begin{align}
P_{\rm{H}}&\equiv(\bar{\ve{p}}_{\frac{1}{2}},
\bar{\ve{p}}_{\frac{3}{2}},\cdots),\\
Q_{\rm{I}}&\equiv(\bar{\ve{q}}_1,\bar{\ve{q}}_2,\cdots),\\
\Lambda_{\rm{H}}&\equiv\mbox{diag}(\kappa_{n-\frac{1}{2}}),\\
\Lambda_{\rm{I}}&\equiv\mbox{diag}(\kappa_{n}),
\end{align}
with
\begin{align}
&\K^2\bar{\ve{p}}_{n-\half}=\kappa_{n-\half}^2\bar{\ve{p}}_{n-\half},\\
&\K^2\bar{\ve{q}}_{n}=\kappa_{n}^2\bar{\ve{q}}_{n},\\
&C\bar{\ve{p}}_{n-\half}=-\bar{\ve{p}}_{n-\half},\\
&C\bar{\ve{q}}_{n}=\bar{\ve{q}}_{n}.
\end{align}
The bar on the vectors represents that they are normalized.

Before evaluating the divergent factor $D$, we must fix the twist-odd
vector $\ve{h}$.
We adopt as $E\ve{h}$ any one of the twist-odd eigenvectors 
$\bar{\ve{p}}_{n-\half}$, which constitute an orthonormal basis of the
twist-odd subspace in the $L$ dimensional vector space.
Using the fact that $(\K\uu)\vL=(1,0,0,\cdots,0)$ and that the first
component of $\bar{\ve{p}}_{n-\half}$ is given by \cite{Hata:2002it}
\begin{equation}
[\bar{\ve{p}}_{n-\half}]_1=\sqrt{\frac{2\pi\kappa_{n-\half}}
{\sinh\bigl(\kappa_{n-\half}\pi/2\bigr)\ln L}},
\end{equation}
$D$ is evaluated as
\begin{align}
D&=-\half\sum_m(\bar{\ve{p}}_{n-\half}\cdot\bar{\ve{p}}_{m-\half})
\frac{1}{\kappa_{m-\half}}[\bar{\ve{p}}_{m-\half}]_1 \nn\\
&=-\half\frac{1}{\kappa_{n-\half}}[\bar{\ve{p}}_{n-\half}]_1 \nn\\
&=-\half\sqrt{\frac{2\pi}
{\kappa_{n-\half}\sinh\bigl(\kappa_{n-\half}\pi/2\bigr)\ln L}}.
\end{align}
For $n\ll\ln L$, the eigenvalue $\kappa_{n-\half}$ is given by
\begin{equation}
\kappa_{n-\half}=\frac{2\pi}{\ln L}\left(n-\half\right),
\end{equation}
and for such $n$, we have
\begin{equation}
D=-\frac{\sqrt{\ln L}}{2\pi(n-\half)}.
\label{DeqsqrtlnL/n}
\end{equation}
On the other hand, when $n$ is of the same order or larger than
$\ln L$, $D$ is much smaller than $\sqrt{\ln L}$ and is insufficient
to compensate the infinitesimal factor
$\varepsilon\sim1/\sqrt{\ln L}$.
Because the eigenvectors $\bar{\ve{p}}_{n-\half}$ for $n\ll\ln L$
approach the zero eigenvectors $\bar\ff^{(0)}$ in the limit
$L\rightarrow\infty$, we conclude that $D$ generates the gauge
transformation only if $E\ve{h}$ is proportional to
$\ff^{(0)}$.

\section{The inverse of $\K^2\vL$}
In the calculation in (\ref{T*t odd}), we must evaluate the quantity
$\K\vL(\K^2\vL)^{-1}\K\vL$. For this purpose, we present the explicit
expression of the inverse of $\K^2\vL$.
Here, we consider the case of $L=\mbox{odd}$.
First, recall that the $L$-dimensional cutoff of the vectors
$\ff^{(0)}$ (\ref{fzero}) and $\uu$ (\ref{u}) is given by
\begin{align}
\ff^{(0)}\vL&=
\left(
1,0,-\frac{1}{\sqrt{3}},0,\frac{1}{\sqrt{5}},0,
\cdots,0,\frac{\sign{L-1}}{\sqrt{L}}
\right)^\T
,\\
\uu\vL&=
\left(
0,-\frac{1}{\sqrt{2}},0,\frac{1}{\sqrt{4}},0,
\cdots,0,\frac{\sign{L-1}}{\sqrt{L-1}},0
\right)^\T
.
\end{align}
In addition to these vectors, we define $\ff_n$ and $\uu_n$ which are
the "truncated" vectors of $\ff^{(0)}\vL$ and $\uu\vL$:
\begin{equation}
\ff_{n}
=\left(
1,0,-\frac{1}{\sqrt{3}},0,\frac{1}{\sqrt{5}},0,
\cdots,0,\frac{\sign{n-3}}{\sqrt{n-2}},
0,\frac{\sign{n-1}}{\sqrt{n}},0,0,\cdots,0,0
\right)^\T,
\end{equation}
for odd $n$ and
\begin{equation}
\uu_{n}
=\left(
0,0,
\cdots,0,0,\frac{\sign{n}}{\sqrt{n}},0,
\frac{\sign{n+2}}{\sqrt{n+2}},0,
\cdots,0,\frac{\sign{L-1}}{\sqrt{L-1}},0
\right)^\T,
\end{equation}
for even $n$.
Using the explicit representation of $\K\vL$ and $\K^2\vL$,
\begin{align}
\K\vL&=
\Smatrix{
0 & -\sqrt{1\cdot 2}
\\
-\sqrt{1\cdot 2} & 0 & -\sqrt{2\cdot 3}
\\
& -\sqrt{2\cdot 3} & 0 & -\sqrt{3\cdot 4}
\\
& & -\sqrt{3\cdot 4} & 0 &\ddots
\\
& & & \ddots & 
\\
& & & & -\sqrt{(L-2)(L-1)} & 0 & -\sqrt{(L-1)L}
\\
& & & & & -\sqrt{(L-1)L} & 0
} ,
\\
\K^2\vL&=
\Smatrix{
2\cdot 1^2 & 0 & 2\sqrt{1\cdot 3}
\\
0 & 2\cdot 2^2 & 0 & 3\sqrt{2\cdot 4}
\\
2\sqrt{1\cdot 3} & 0 & 2\cdot 3^2 & 0 & \ddots
\\
& 3\sqrt{2\cdot 4} & 0 & \ddots
\\
& & \ddots & & & \ddots
\\
& & & & 2(L-2)^2 & 0 & (L-1)\sqrt{L(L-2)}
\\
& & & \ddots  & 0 & 2(L-1)^2 & 0
\\
& & & & (L-1)\sqrt{L(L-2)} & 0 & 2 L^2
} ,
\end{align}
We can calculate the action of $\K\vL$ and $\K^2\vL$ on the truncated
vectors:
\begin{align}
\K\vL\ff_n&=-\sign{n-1}\en{n+1}, \label{product-Kf}\\
\K\vL\uu_n&=-\sign{n}\en{n-1} -\sign{L-1}\en{L}, \label{product-Ku}\\
\K^2\vL\ff_n&=\sign{n-1}(n+1)
\left(\en{n}+\en{n+2}\right),\\
\K^2\vL\uu_n&=\sign{n}(n-1)
\left(\en{n}+\en{n-2}\right)+\sign{L-1}L\en{L-1}
\label{product-K2u}
\end{align}
where $\ee_n$ is the unit vector along the $n$-th direction; 
$[\ee_n]_m=\delta_{nm}$.
In (\ref{product-Kf})--(\ref{product-K2u}), we have $\ee_n=0$ when
$n>L$ and $n<1$.
Using these equations, we can show that the inverse of $\K^2|_L$
is expressed as follows:
\begin{equation}
\label{K of d}
(\K^2\vL)^{-1}=(\dd_1,\dd_2,\dd_3,\cdots,\dd_{L-1},\dd_L)
\end{equation}
where the vector $\dd_n$ is defined by
\begin{align}
\dd_{2k+1}&=\frac{(-1)^{k}}{\sqrt{2k+1}}
\sum_{n=k+1}^{(L+1)/2}\frac{1}{2n}\ff_{2n-1} ,\\
\dd_{2k}&=\frac{(-1)^{k}}{\sqrt{2k}}\left(
\sum_{n=1}^{k}\frac{\uu_{2n}}{2n-1}
-\frac{\sum_{m=1}^{k}\frac{1}{2m-1}}{
\sum_{m=1}^{(L+1)/2}\frac{1}{2m-1}}
\sum_{n=1}^{(L-1)/2}\frac{1}{2n-1}\uu_{2n}
\right).
\end{align}
{}From (\ref{product-Kf})--(\ref{product-K2u}), 
(\ref{K of d}), we finally obtain
\begin{align}
\left[{\bf{1}}_L -\K\vL\frac{1}{\K^2\vL}\K\vL\right]_{n,m}
&=\left(\sum_{k=1}^{[L/2]+1}\frac{1}{2k-1}\right)^{-1}
[\ff^{(0)}\vL]_n [\ff^{(0)}\vL]_m\nn\\
&=[\bar{\ff}^{(0)}\vL]_n [\bar{\ff}^{(0)}\vL]_m\label{ff}.
\end{align}
This is the projection operator which singles out the zero-mode of
$\K$. Eq.\ (\ref{ff}) holds also in the case $L=\mbox{even}$.

\end{document}